\documentclass[twocolumn,amsmath,amsfonts,amssymb,showpacs]{revtex4} 
\usepackage{graphicx}
\def\beq{\begin{equation}}
\def\eeq{\end{equation}}
\def\bea{\begin{eqnarray}}
\def\eea{\end{eqnarray}}

\begin{document}

\title{Infinite Statistics Condensate as a Model of Dark Matter}

\author{Zahra Ebadi}
\email{z.ebadi@ph.iut.ac.ir}
\affiliation{Department of Physics, Isfahan University of Technology, Isfahan 84156-83111, Iran}
\author{Behrouz Mirza}
\email{b.mirza@cc.iut.ac.ir}
\affiliation{Department of Physics, Isfahan University of Technology, Isfahan 84156-83111, Iran}
\author{Hosein Mohammadzadeh}
\email{mohammadzadeh@uma.ac.ir}
\affiliation{Department of Physics, University of Mohaghegh Ardabili, P. O. Box 179, Ardabil, Iran}

\begin{abstract}
In some models, dark matter is considered as a condensate  bosonic system. In this paper, we prove that condensation is also possible for particles  that obey infinite statistics and derive the critical condensation temperature. We argue that a condensed  state  of a gas of very weakly interacting particles obeying infinite statistics could be considered as a consistent model of dark matter.

\end{abstract}

\pacs{05.30.-d, 67.85.Jk, 95.35.+d,  04.60.-m}
\maketitle

\section{Introduction}
 Investigation of cosmological nucleosynthesis leads to the conclusion that most of the mass in the universe is not in the form of ordinary baryonic matter, i.e. atomic nuclei and electrons. This conclusion is strongly supported by the  anisotropies observed in the cosmic microwave background \cite{winberg,lyth}.
 These observations  reveal a universe composed of $4$\% baryons, $22$\% non-baryonic dark matter, and $74$\% dark energy \cite{lyth,jarosik}.

 Dark matter plays a key role in the current models of galaxy formation \cite{winberg}, but one of the most important questions in  modern cosmology is what the dark matter is made up of.
  Observation of  rotation curves \cite{ben2}, gravitational lensing \cite{ben3}, and X-ray spectra \cite{ben4} are useful for determining the mass distribution of dark matter. The data thus obtained have led to the conclusion that galaxies are composed of a luminous galactic disk surrounded by a spherical galactic halo of dark matter which comprises roughly $95$\% of the total mass of the galaxy \cite{ben5}. We expect  the constituent particles of dark matter to have no charge, and to be cold and long-lived \cite{winberg}.
Various models have been proposed to explain dark matter. Some maintain that dark matter is made up of
 Bose-Einstein condensate \cite{pathria} particles with a very small mass \cite{sin,hu,arbey,Harko7,harko}. The condensation process
 is interpreted as a phase transition taking place sometime during the cosmic history of the universe. As the normal bosonic dark matter cools below the critical temperature $T_{cr}$ of condensation, it becomes energetically favorable to forming a condensate in which all the particles are in the same quantum state. Different properties of the Bose-Einstein condensed dark matter have been investigated  \cite{fuku,kain,gonzales}.

  The standard approach to treating a nonrelativistic Bose-Einstein condensation is to use the Gross-Pitaevkii equation \cite{gross}, also known as  the nonlinear Schr\"{o}ndinger equation which is in the form of the Schr\"{o}ndinger equation with an interaction term and trap potential introduced.  There is a constant coupling between the interacting term and the Newtonian gravitational potential which is proportional to the scattering length of two interacting bosons and the trap potential, so that the Gross-Pitaevskii equation is coupled with the Poisson equation \cite{harko,kain,wang,brook}. Using this equation, one can achieve the basic astrophysical properties of the condensate, i.e. density profile, rotational velocity, and mass profile  \cite{Harko7,harko11}. Dwarf galaxies have been already studied \cite{harko11}, these are dark matter dominated astrophysical objects with a very small contribution of baryons to the total matter content, which make them suitable for testing models of dark matter. Recently, the intermediate statistics is used in various aspects of gravitation and cosmology \citep{MirzaC,Rafaelli}.

In  models  based on the Bose-Einstein condensation, constituent particles of dark matter are considered to be bosons. However, theories based on the holographic model of spacetime foam do not regard them as bosons \cite{Ng}. Some physicists believe that space is composed of an ever-changing arrangement of bubbles called spacetime foam, also known as quantum foam \cite{Ng,spacetime foam}. This model necessitates that the dark energy particles obey not the ordinary Bose or Fermi statistics but the exotic infinite statistics
 \cite{Ng}. Infinite statistics \cite{greenberg1,greenberg90,relativity} is defined by  generalizing  commutation and anticommutation  relations between the creation and annihilation operators.

 The present paper is organized as follows. In Sec. II, we review infinite statistics. In Sec.  III, we summarize the reasons why infinite statistics must be employed in dealing with dark matter. We will present some arguments about the infinit statistics and cold dark matter in section IV. In Sec. V, we investigate the condensation and critical temperature of a system with particles obeying infinite statistics and compare this approach with the well-known Bose-Einstein condensation.  We also consider the condensation of particles obeying infinite statistics as a model for dark matter. Hence, we will summarize below the statistics of quons in order to work out the critical temperature of condensation.

\section{Quons and Infinite Statistics }

The creation and annihilation operators of bosons and fermions satisfy the well known commutation and anticommutation algebra, $a_{i}a_{j}^{\dag}\mp a_{j}^{\dag}a_{i}=\delta_{ij}$. This relation requires  that all particles be in either a symmetric  or an antisymmetric state. Without such a symmetrization postulate, there are many generalizations that may have intermediate properties of Bose and Fermi statistics, namely para-Bose,  para-Fermi, $q$-deform bosons and fermions, $q$-mutator and quons \cite{greenberg90,greenberg1,govorkov,greenberg2}.
 Quons are characterized by a $q$-deformation of the commutation relations:
 \bea
a_{i}a_{j}^{\dag}-qa_{j}^{\dag}a_{i}=\delta_{ij},
\eea
 $q=\pm1$ corresponds to bosons or fermions, while other cases ($-1<q<1$) correspond to quons. The special case $q=0$ is called infinite statistics.
 ,Therefore it is characterized by
\bea
a_{i}a_{j}^{\dag}=\delta_{ij}
\eea
 and a unique vacuum state annihilated by all the annihilators $a_{i}$:
 \bea
 a_{i}|0\rangle=0.
 \eea
This relation determines a Fock-state representation in a linear vector space. States are built by acting on a vacuum. According to Eq. (2), there is no commutation relation between two annihilation or creation operators for infinite statistics. The quantum states are orthogonal under any permutation of the identical particles and, therefore, all representations of the symmetric group will occur. Hence, infinite statistics can be interpreted as the statistics of identical particles with an infinite number of internal degrees of freedom, which is equivalent to the statistics of nonidentical particles since they are distinguishable by their internal states \cite{Ng,relativity}.
 The loss of local commutativity implies violation of locality, which is an important character of quantum gravity. By virtue of these properties, infinite statistics has been applied in a variety of areas, such as black hole statistics \cite{strominger,volovich}, dark energy quanta \cite{Ng,ng}, large $N$ matrix theory \cite{douglas}, and holography principle \cite{shevchenko}.
The statistical mechanics of particles obeying infinite statistics can be derived in a  way similar to the usual derivation of Boltzmann statistics. The partition function $Z_{N}$ is given by
\bea
Z_{N}= \sum_{quantum\  states}e^{-\beta H}
\eea
for a given set of occupation numbers $n(p)$, where $p$ is a momentum, with $N=\sum_{p}\ n(p)$, there are
\bea
g[n(p)]=N!/\prod_{p}n(p)!
\eea
orthogonal quantum states \cite{greenberg90}. This is just Boltzmann counting without the Gibbs factor$1/N!$.

M. V. Medvedev \cite{medvedev} introduced an ambiguous statistics which may exhibit both Bose and Fermi statistics with the respective probabilities $p_{b}$ and $p_{f}$.
One can consider a realization of a system with $N_{j}$ particles where in each realization, the system has $k$ bosons and $N_{j}-k$ fermions while the probability of this realization is $p_{b}^{k}p_{f}^{N_{j}-k}$. Therefore, the total number of states of the system will be
 \bea
 W=\prod_{j}\sum_{k=0}^{N_{j}}\frac{N_{j}!}{k!(N_{j}-k)!} w_{bj}(k)w_{fj}(N_{j}-k)p_{b}^{k}p_{f}^{N_{j}-k},
 \eea
 where, $w_{b}(m)$ and $w_{f}(m)$ are the number of quantum states of $n$ identical particles occupying a group of $G$ states for bosons and fermions, respectively:
 \bea\nonumber
 w_{b}(m)=\frac{(G+m-1)!}{m!(G-1)!}  ,\   w_{f}(m)=\frac{G!}{m!(G-m)!}
 \eea

Using the maximum entropy for the most probable distribution and following the technique of Lagrange's multipliers, the occupation number is given by
 \bea
 n_{j}=\frac{\sigma\rho}{(\rho+p_{f})((\rho-p_{b})}[1+\sqrt{1-\frac{\delta^{2}(\rho+p_{f})((\rho-p_{b})}{\sigma^{2}\rho(\rho+\delta)}}], \
 \eea
where, $\delta=p_{f}-p_{b}$, $\sigma=p_{f}+p_{b}$, $\rho=\exp{(\epsilon_{j}-\mu)/k_{B}T}$ and $\mu$ is the chemical potential \cite{medvedev}.

Also, it has been shown that there is a conjectured relation between the particles of the ambiguous statistical type and the quons \cite{medvedev}. It was argued that  $q$ in Eq. (1) is related to the parameters of ambiguous statistics as follows
 \bea
 q=\frac{p_{b}-p_{f}}{p_{b}+p_{f}}
 \eea
  One can find the distribution function of infinite statistics ($q=0$) as follows,
  \bea
  n(\epsilon)=\frac{4p\rho}{\rho^{2}-p^{2}}
  \eea
where, $p=p_{f}=p_{b}$. Therefore, the thermodynamic quantities such as the internal energy and the total number of particles for an ideal gas with infinite statistic gas will be
 \bea
 U&=&\int_{0}^{\infty}\epsilon \
 n(\epsilon)\Omega(\epsilon)d\epsilon=\frac{A}{\beta^{d/2+1}}\int_{0}^{\infty}\frac{4pze^{x}x^{d/2}}{e^{2x}-p^{2}z^{2}}dx\nonumber\\
 N&=&\int_{0}^{\infty}n(\epsilon) \Omega(\epsilon)d\epsilon=\frac{A}{\beta^{d/2}}\int_{0}^{\infty}\frac{4pze^{x}x^{d/2-1}}{e^{2x}-p^{2}z^{2}}dx\label{pn}
 \eea
 where, $\Omega(\epsilon)=A\epsilon^{d/2 -1}=\frac{V}{\Gamma(\frac{d}{2})}\frac{(2m\pi)^{d/2}}{h^{d}}\epsilon^{d/2 -1}$ is the density of the single particle state, $V$ is the volume of the system, $z=\exp(\beta\mu)$ is the fugacity, and $x=\beta\epsilon$.

\section{Infinite Statics and Holographic Foam Model of Spacetime }

In this section, we summarize the reason  one has to use infinite statistics for dark energy.
Quantum fluctuations of spacetime give rise to quantum foam, and black hole physics dictates that the foam is of holographic type \cite{Ng}.
Applied to cosmology, the holographic model requires the existence of a specific dark energy that is composed of an enormous number of inert particles of extremely long wavelengths \cite{Ng}.
The entropy of an ideal gas is given by
 \bea
 S=Nk_{B}[\ln(\frac{V}{N\lambda^{3}})+5/2].\label{entropy}
 \eea
where, $N$ is the total number of particles in a volume $V$ and  $\lambda=(2\pi\hbar^{2}/mk_{B}T )^{1/2}$ is thermal wavelength\cite{pathria}. Y. Ng has employed this equation for a gas of  dark energy \cite{ng13}. In the above equation, if $V\thicksim\lambda^{3}$ (long wavelength), the entropy will be negative unless $N\thicksim1$ or if the Gibbs factor $1/N!$ in the partition function is removed from the calculation of the entropy. The first condition is not possible because the holographic foam model of spacetime  has been used to show that dark energy is a cosmological manifestation of quantum foam and that it consists of a large number of very low energy particles \cite{Ng}. According to the holographic foam model, $N$ should not be too different from $(R_{H}/l_{P})^{2}\gg1$ \cite{Ng,ng17}, where $R_{H}$ and $l_{P}$ are the Hubble radius and the planck length, respectively. Therefore,  the alternative condition will be feasible which will lead  to the following equation:

\bea
 S=Nk_{B}[\ln(\frac{V}{\lambda^{3}})+3/2].
 \eea

 Removing the Gibbs factor shows that these particles are neither fermions nor bosons and that they obey some other statistics \cite{ng13}. The only known consistent statistics in $D>2$ without the Gibbs factor is infinite statistics or quantum Boltzmann statistics \cite{ng14,greenberg90}.
\section{Infinite Statistics and Dark Matter}
If we use the Eq. (\ref{entropy})  for cold dark matter at very low temperature, the negative entropy problem will be occurred. In other words, in very low temperature limit $N\lambda^{3}/V\gg1$ and the first term in Eq. (\ref{entropy}) will be negative. We argue that in this limit to avoid the negative entropy, we need a statistics with removed Gibbs factor. There are some models based on the condensation of the boson gas for cold dark matter \cite{fuku,kain,gonzales} and therefore, we could propose the condensate of infinite statistics as a model of dark matter. In comparison with the Y. Ng model for dark energy, we notice that $V$ is not in order of $\lambda^{3}$ but the temperature is low enough. Therefore, we could employ the the infinite statistic for dark matter in low temperature limit. We expect that the condensation in low temperature for ideal boson gas and we explore the possibility of condensation for infinite statistics in next section. Also, we will show that infinite statistics condensate (ISC) model for dark matter may be lead to less interacting dark matter particles with respect to the BEC models.
Recently, It has been shown that the unconventional quanta of MONDian dark matter must obey infinite statistics which contains some similar ideas as this paper\cite{Ho}.

\section{Infinite Statics and Condensation }

\begin{figure}[b]
    \center
    \includegraphics[width=0.9\columnwidth]{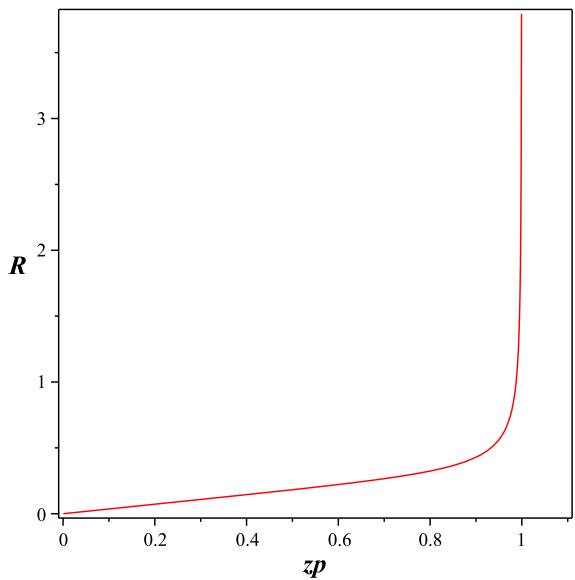}\\
    \caption{(Color online) Thermodynamic curvature of an ideal gas with particles obeying infinite statistics with respect to the fugacity of gas for isotherm processes ($\beta=1$). }\label{fig}
   \end{figure}

 Based on the observations in the preceding section, we explore the condensation of particles obeying infinite statistics as a model for dark matter and work out the critical condensation temperature for this statistics. Investigation of the thermodynamic geometry of some thermodynamic systems has led to the introduction of a qualitative tool, namely, the thermodynamic curvature that is singular at the phase transition point and its sign specifies the statistical interaction of the system \cite{Ruppeiner1,Ruppeiner01,Ruppeiner2010,Mrugala2,Mirza}. We construct the thermodynamic geometry of an ideal gas with particles obeying infinite statistics. The metric elements of the thermodynamic parameter space using the Fisher-Rao representation are given by
 \bea
 \label{vv} &&G_{\beta\beta}=\frac{\partial^{2}\ln\cal{Z}}{\partial
    \beta^{2}}=-(\frac{\partial U}{\partial\beta})_{\gamma},\nonumber\\
    &&G_{\beta\gamma}=G_{\gamma\beta}=\frac{\partial^{2}\ln\cal{Z}}{\partial\gamma\partial\beta}=-(\frac{\partial N}{\partial \beta})_{\gamma},\\
     &&G_{\gamma\gamma}=\frac{\partial^{2}\ln\cal{Z}}{\partial\gamma^{2}}=-(\frac{\partial N}{\partial\gamma})_{\beta},\nonumber
    \eea
 where, $\beta=1/k_{B}T$ and $\gamma=-\mu/k_{B}T$.
 The curvature of two dimensional thermodynamic parameter space can be evaluated by the Christoffel symbol,
$\Gamma_{ijk}=\frac{1}{2}\left(G_{ij,k}+G_{ik,j}-G_{jk,i}\right)$,
using the derivative of the metric elements with respect to the
thermodynamic parameters. In the following, one can evaluate the
well-known Riemann tensor, Ricci tensor and, finally,  Ricci
scalar which will be called the thermodynamic curvature because
of the identity of the constructed geometry. The thermodynamic curvature of an ideal gas with particles obeying infinite statistics in a three dimensional box as a function of fugacity for isothermal processes has been depicted in Fig. \ref{fig}.
 Clearly, the thermodynamic curvature is positive and the statistical interaction is, therefore, attractive.
Also, there is a critical point $(z=1/p)$ where the thermodynamic curvature is singular. This is similar to the ideal boson gas wherein the statistical interaction is positive and the thermodynamic curvature is singular at the Bose-Einstein condensation point ($z=1$).
  Bose-Einstein condensation occurs at the singular point of the occupation number at $\epsilon=0$ and $\mu=0$. According to Eq. (10), within the infinite statistics limit,  condensation occurs at $z=1/p$ or equivalently at,

 \bea
 \mu=-k_{B}T \ln p.
 \eea
 Now, we can evaluate the phase transition temperature using the total number of particles obeying infinite statistics. For non-relativistic particles with the dispersion relation
 $\epsilon(k)={\hbar^{2}k^{2}}/{2m}$ and at $z =1/p$, according to  Eq. (\ref{pn}) we will have
   \bea
   n=\frac{N}{V}=\frac{2(\pi)^{d/2}}{(2\pi)^{d}\Gamma(d/2)}\int_{0}^{\infty}dk k^{d-1}\frac{4\exp(\epsilon/k_{B}T_{cr})}{\exp(2\epsilon/k_{B}T_{cr})-1}\ \label{temperature}
   \eea
 where, $k$ is the wave number.
 The finite critical temperature for particles obeying infinite statistics  can be compared to Bose-Einstein condensation temperature.
  Solving  Eq. (\ref{temperature}) for the critical temperature yields,
     \bea
     k_{B}T_{cr}=\frac{2\pi \hbar^{2}}{m_{q=0}}(\frac{n}{\zeta(d/2)})^{2/d}\frac{2^{1-4/d}}{(2^{d/2}-1)^{2/d}},\label{temperature1}
     \eea
 where, $m_{q=0}$ is the quon mass that constitutes the dark matter and $\zeta(x)$ is the Riemann zeta function analytically continued to all $x\neq1$. Also,  Eq. (\ref{temperature1}) indicates that there is a finite critical temperature for $d>2$. We can obtain some properties of quon condensation by comparison with a boson gas condensation. For a bosonic system, the critical temperature of Bose-Einstein condensation is given by \cite{pathria}
   \bea
     k_{B}T_{cr}=\frac{2\pi \hbar^{2}}{m_{b}}(\frac{n}{\zeta(d/2)})^{2/d}.\label{temperature2}
     \eea
  where, $m_{b}$ is the boson mass that constitutes the dark matter.
    One can compare the critical temperature in Eq. (\ref{temperature1}) with that of Bose-Einstein condensation given by Eq. (\ref{temperature2}) to find that the critical temperature of the infinite statistics model has a correction factor unlike the Bose-Einstein condensation temperature
    \bea
    \frac{2^{1-4/d}}{(2^{d/2}-1)^{2/d}}.\label{factor}
    \eea
  The critical temperatures of quon and boson particles, Eqs. (\ref{temperature1}, \ref{temperature2}), with a specified value of density number will be the same if the relation between $m_{b}$ and $m_{q=0}$ is of the following form
  \bea
\frac{m_{q=0}}{m_{b}}= \frac{2^{1-4/d}}{(2^{d/2}-1)^{2/d}}.\label{mm}
\eea
We obtain $m_{q=0}\simeq0.53 m_{b}$ for $d=3$.
It has been shown that the current energy density of the universe is $\rho_{cr}=9.44\times 10^{-30} g/cm^{3}$ and so Bose-Einstein condensation takes place provided that the boson mass satisfies the restriction $m_{b}<1.87 eV $ \cite{fuku}. It is easy to find a similar limit for the mass of quons.
So we propose that a condense state of quon particles that obey infinite statistics is a new good candidate for investigating the dark matter.

One can also consider $m_{b}= m_{q=0}$ and according to Eqs. (\ref{temperature1}) and (\ref{temperature2}), the critical temperature of infinite statistics model will be smaller than Bose-Einstein condensation temperature. It is known that the self interaction
of dark matter particles must be very weak.  This is because of the observations of elongated shapes of clusters of galaxies \cite{Ling}. In the following we will show that our model represents a condensed state of very weakly interacting particles which is almost similar to non-interacting particles of an ideal gas.
   When the condensation occurs, we can work out the equation of state. It has been shown that the equation of state for an ideal gas of particles obeying the infinite statistics is given by \cite{medvedev}
\bea
PV&=&\frac{2}{3}\int_{0}^{\infty}\epsilon \
 n(\epsilon)\Omega(\epsilon)d\epsilon\nonumber\\
 &=&V\frac{(2m\pi)^{3/2}}{h^{3}}\frac{2^{5/2}-1}{2^{1/2}}\zeta(5/2)(k_{B}T)^{5/2}.
\eea
Using Eqs. (\ref{pn}) and (\ref{temperature1}) at the critical point, we obtain
\bea
PV=\alpha Nk_{B}T_{cr}
\eea
where, $\alpha=\frac{\zeta(5/2)}{2\zeta(3/2)}\frac{2^{5/2}-1}{2^{3/2}-1}\simeq0.65$ is a constant coefficient. Roughly the equation of state is $PV\simeq Nk_{B}T$ ; which means that the particles effectively behave as an ideal  gas (very weakly interacting) in the condensate regime. The related condensation  temperature is also lower than the Bose-Einstein condensation temperature, which means a lower  pressure for the gas.
At low temperature in the condensate regime, the thermal wavelength is very  long, which indicate that the particles or the physical degrees of freedom are highly correlated.  Although, the system is correlated, but the internal interaction of particles is very weak.

Therefore, condensed state of  a gas of very weakly interacting particles obeying infinite statistics presents a model which indicates that the level of self interaction is small enough to produce effectively collisionless dark matter particles.



\section{Conclusion and Discussion }

Different models have been proposed to describe dark matter. Some of these models are based on the Bose-Einstein condensate and are consistent with such observational data as density profile of dwarf galaxies \cite{Harko7}.  We argued that constituent particles of dark matter could obey infinite statics rather than Bose or Fermi statistics. Therefore, we investigated the possibility of quon condensation using infinite statistics.  We introduced the first  model of a condensed state of a  gas of very weakly interacting particles obeying infinite statistics.

According to the thermodynamic curvature of the thermodynamic parameter space, the ISC has an attractive internal intrinsic interaction and at the critical point the equation of state is $PV\simeq Nk_{B}T$, which describes an ideal  gas with non-interacting particles. Therefore, it is obvious that the pressure will go to zero at the condensate regime where the temperature is very low. Dark matter is thought to be cold and nearly pressureless \cite{Ling}.
Therefore, the presented model again is in a good consistency with the standard cold dark matter.






\begin{thebibliography}{99}

\bibitem{winberg}
S. Weinberg, {\it Cosmology}, Oxford University Press (2008).


\bibitem{lyth}
D. Lyth and A. Liddle, {\it The Primordial Density Perturbation}, Cambridge University Press (2009).


\bibitem{jarosik}
N. Jarosik {\it et al.}, Astrophys. \ J. \ Suppl. \ {\bf 192}, 14 (2011)


\bibitem{ben2}
D. A. Dale, R. Giovanelli, M. P. Haynes, E. Hardy and L. Campusano. AJ, 115, 418 (1998);
 V. C. Rubin, N. Thonnard and W. K. J. Ford, Astrophys. J. {\bf 238}, 471 (1980).


\bibitem{ben3}
 A. Refregier, Ann. Rev. Astron. Astrophys. {\bf 41}, 645 (2003).


\bibitem{ben4}
A. D. Lewis, D. A. Buote and J. T. Stocke, astrophys. J. {\bf 586}, 135 (2003).


\bibitem{ben5}
W. H. Press, B. S. Ryden and D. N. Spergel, Phys. Rev. Lett. {\bf 64}, 1084 (1990);
K. Freese, arXiv:astro-ph/0812.4005

\bibitem{pathria}
R. K. Pathria, {\it Statistical mechanics}, 2nd ed. Elsevier Ltd. (2000).

\bibitem{sin}
S. J. Sin, Phys. Rev. D {\bf 50}, 3650 (1994).

\bibitem{hu}
W. Hu, R. Barkana and A. Gruzinov, Phys. Rev. Lett. {\bf 85},1158(2000)

\bibitem{arbey}
 A. Arbey, J. Lesgourgues, P. Salati, Phys. Rev. D {\bf 64}, 123528 (2001).

\bibitem{harko}
C. G. Boehmer and T. Harko, JCAP {\bf 0706}, 025 (2007).

\bibitem{Harko7}
 T. Harko, Phys. Rev. D {\bf83}, 123515 (2011).

\bibitem{fuku}
T. Fukuyama, M. Morikawa and T. Tatekawa, JCAP {\bf 0806}, 033 (2008).

\bibitem{kain}
B.Kain and H. Y. Ling, Phys. Rev. D {\bf 82}, 064042 (2010).

\bibitem{gonzales}
J. A. Gonzalez and F. S. Guzman, Phys. Rev. D {\bf 83} 103513 (2011).

\bibitem{gross}
F. Dalfovo,S. Giorgini, L. P. Pitaevskii and S. Stringari, Rev. Mod. Phys. {\bf 71}, 463 (1999); C. J. Pethick and H. Smit,
{\it Bose-Einstein condensation in dilute gases}, Cambridge, Cambridge University Press, (2008).

\bibitem{wang}
X. Z. Wang, Phys.Rev. D {\bf 64}, 124009 (2001).

\bibitem{brook}
M. N. Brook and P. Coles, arXiv:0902.0605 [astro-ph.Co].

\bibitem{harko11}
T. Harko, JCAP {\bf 05}, 022 (2011).

\bibitem{MirzaC}
S. Zare, Z. Raissi, H. Mohammadzadeh and B. Mirza, Euro. phys. J. C {\bf 72} 2152 (2010).

\bibitem{Rafaelli}
B. Rafaelli, JHEP {\bf 01,} 188 (2013).

\bibitem{Ng}
Y. J. Ng, arXiv:1001.0411[gr-qc] (2010)


\bibitem{spacetime foam}
L. J. Garay, Int. J. Mod. Phys. A {\bf 10}, 145-146 (1995);
S. Hossenfelder, Mod. Phys. Lett, A {\bf 19}, 2727-2744 (2004).


\bibitem{greenberg1}
H. S. Green, Phys. Rev. {\bf 90}, 270 (1953)

\bibitem{greenberg90}
O. W. Greenberg, Phys. Rev. Lett. {\bf 64}, 705 (1990)

\bibitem{relativity}
Chao Cao, Yi-Xin Chen and Jian-Long Li, Phys. Rev. D {\bf 80}, 125019 (2009)

\bibitem{govorkov}
A. B. Govorkov, Phys. Part. Nucl. {\bf24}, 565 (1993)

\bibitem{greenberg2}
O. W. Greenberg, Phys. Rev. D {\bf 43}, 4111 (1991)


\bibitem{strominger}
A. Strominger, Phys. Rev. Lett. {\bf 71}, 3397 (1993).

\bibitem{volovich}
I. V. Volovich, arXiv:hep-th/9712202

\bibitem{ng}
Y. J. Ng, Phys. Lett. B {\bf 657}, 10 (2007); V. Jejjala, M. Kavic and D. Minic, Adv. High Energy Phys. {\bf 2007}, 1 (2007); M. Li, X. D. Li, C. S. Lin and Y. Wang, Commun. Theor. Phys. {\bf 51}, 181 (2009); A. J. M. Medved, Gen. Relativ. Gravit. {\bf 41}, 287 (2009).

\bibitem{douglas}
M. R. Douglas, Phys. Lett. B {\bf 344}, 117 (1995); R. Gopakumar and D. Gross, Nucl. Phys. {\bf b451}, 379 (1995).

\bibitem{shevchenko}
V. Shevchenko, Mod. Phys. Lett. A {\bf 24}, 1425 (2009).

\bibitem{medvedev}
M. V. Medvedev, Phys. Rev. Lett. {\bf 78}, 4147 (1997).

\bibitem{ng17}
M. Arzano, T. W. Kephart and Y. J. Ng, Phys. Lett. B {\bf 649}, (2007) 243; M . Maziashvilli, arXive:gr-qc/0612110

\bibitem{ng13}
Y. J. Ng, Phys. Lett. B {\bf 657}, 10 (2007)


\bibitem{ng14}
S. Doplicher, R. Haag and J. Roberts, Commun. Math. Phys. {\bf 23}, 199(1971)

\bibitem{Ho}
C. M. Ho, D. Minic and Y. J. Ng, Phys. Rev. D {\bf 85,} 104033 (2012).

\bibitem{Ruppeiner1}
G. Ruppeiner, Phys. Rev. A {\bf 20}, 1608 (1979).

\bibitem{Ruppeiner01}
G. Ruppeiner, Rev. Mod. Phys. {\bf 67}, 605 (1995).

\bibitem{Ruppeiner2010}
G. Ruppeiner, Am. J. Phys. {\bf 78}, 1170 (2010).

\bibitem{Mrugala2}
H. Janyszek and R. Mruga{\l}a, J. Phys. A: Math. Gen. {\bf 23},
467 (1990); H. Janyszek and R. Mruga{\l}a, ibid. {\bf 23},
477 (1990); H. Janyszek and R. Mruga{\l}a, Phys. Rev. A {\bf 39}, 6515 (1989).

\bibitem{Mirza}
B. Mirza and H. Mohammadzadeh, Phys. Rev. E {\bf 82}, 031137 (2010); Phys. Rev. E {\bf 84}, 031114 (2011);
J. Phys. A: Math. Theor. {\bf 44}, 475003 (2011).

\bibitem{Ling}
B. Kain and H. Y. Ling, Phys. Rev. D {\bf 85}, 023527 (2012).



\end{thebibliography}
\end{document}